# A survey on agile practices and challenges of a global software development team

Tatiane Lautert, Adolfo Gustavo Serra Seca Neto, Nádia P. Kozievitch

Universidade Tecnológica Federal do Parana (UTFPR), Curitiba - Brazil

**Abstract.** The Agile Manifesto describes that the most efficient and effective method of conveying information to and within a development team is through face-to-face conversation, however that is not always possible when teams are working in a Global Software Development (GSD) environment. Based on this scenario, this study presents an exploratory data analysis using survey results to explore agile practices and challenges of a global software development team that uses Scaled Agile Framework (SAFe), which is designed to the need of larger organizations. The goal of this study is to understand the team's level of knowledge in some agile practices and which types of communication are usually prioritized. As in GSD environments team members are geographically spread across multiple regions and time zones, we aim to identify challenges this environment can present. As a result of this exploratory analysis, it has been identified that communication is one of main challenges in GSD environment and that phone calls are considered to be the most efficient type of communication. Additionally, we have also identified that professionals have different levels of confidence in Agile practices and concluded that knowledge transfers could help level set teams overall confidence and knowledge.

## 1 Introduction

Agile software development is based on a set of 4 values and 12 principles described in the Agile Manifesto [1]. It was written in 2001 by a group of 17 practitioners interested in finding better ways of developing software that is centered on individuals but also is able to respond to rapid changes. Agile Software development can be described as a lightweight methodology as opposed to heavyweight traditional software engineering processes. One of the principles of the Agile Manifesto describes that the most efficient and effective method of conveying information to and within a development team is through face-to-face conversation, however that is not always possible when teams are working in a Global Software Development (GSD) environment.

According to J. D. Herbsleb and D. Moitra [2], software has become a crucial component for almost every business in recent years and developing software or implementing changes to software that responds to markets' demands is a competitive advantage, vital for business success. Over the recent decades many organizations began to experiment with remotely located software development facilities and with outsourcing, seeking lower costs and skilled resources. The authors highlight that potential benefits of GSD should not be neglected, however a number of problems are also identified and communication is one of them. In order to respond to these rapid market demands, the IT industry has been adopting Agile software development practices and frameworks such as Scrum, Extreme Programming (XP), Lean, Crystal, Dynamic Systems Development Method (DSDM), Feature Driven Development (FDD), and others.

These frameworks or methods provide guidelines which are usually tailored for small teams and serve well for enabling the execution of their development, coordination and communication tasks. However, these methods by themselves do not scale to the need of larger organizations where hundreds of professionals are involved in the development of large and complex solutions [4]. In that scenario, during recent years, several frameworks for scaling agile have been created including Scaled Agile Framework (SAFe), Large-scale Scrum (LeSS) and Disciplined Agile Delivery (DAD) as cited by M. Paasivaara [3].

Scaled Agile Framework (SAFe) [5] was created by Dean Leffingwell and its latest version is SAFe 4.6. It is composed of 4 different configurations, being them: Essential SAFe, Portfolio SAFe, Large Solution SAFe and Full SAFe. Each of these configurations have a set of organization levels (Portfolio, Large Solution, Program and Team) and each level contains details and guidelines about roles, activities, events, and processes applicable to each level. At the Program Level, SAFe uses the concept of the Agile Release Train (ART) which can be described as a virtual organization composed of around 50 to 125 people that are aligned to a business mission and they work together to plan, commit, develop and deploy the solutions. In SAFe's website there is an interactive picture which contains links that takes to web pages with more details on each role, processes, activities, and others that are part of the framework[1].

In this study, a survey was conducted with a GSD team of a large financial services organization that uses SAFe. The team is composed of about 170 professionals that form two Agile Release Trains (ARTs). The professionals are spread across multiple locations, being the majority of them based in the United States, Brazil and India. All members of each Scrum team can be located in the same region or sometimes there could be different configurations as well. Typically, Product Owners, Development Managers, Business Analysts and Program related roles are based in the US while Developers, System Analysts and Technical Leads are based in Brazil or India but that is not a fixed configuration.

The objective of this survey is to understand the level of knowledge in some agile practices of these professionals, which types of communication are usually prioritized and what challenges GSD environments can present. Additionally, the aim is to test the following hypothesis:

[H1] Professionals with more experience in agile methodologies prioritize synchronous or asynchronous communication?

---

[1] https://www.scaledagileframework.com/, last accessed 10-Jun-2019

The data was collected through this survey and in this study an exploratory data analysis is presented.

## 2 Related Work

As highlighted by Hossain et al [8] there is a growing interest in applying agile practices in Global Software Development (GSD) projects. In this paper the authors conduct a systematic literature review of the primary studies that report using Scrum practices in GSD environment and the objective of their study was to identify various challenging factors that restrict the use of Scrum practices in projects that are globally distributed. One of their conclusions is that scrum practices need to be extended or modified in order to support globally distributed software development teams.

Fitriani et al [7] also conducted a systematic literature review and found that there are 30 challenges in implementing Agile Software Development. Among these 30 challenges, the authors concluded that the most significant challenges are team management and distributed team, followed by requirement prioritization, documentation, changing and over-scoping requirement, organization, process, and progress monitoring and feedback.

Other studies that investigate Agile practices and challenges are for example Salinas et al [6] and Nazir el at [9], both papers describe surveys. In the first paper the authors focus on Paraguayan software community and how this community is adopting agile methods. They present initial concerns and barriers of implementation of agile methods in software development companies in Paraguay. In the second paper, the authors focus on the investigation of the extent of agile practices adoption in regards to the Indian IT Industry concluding that agile practices affect the cost and increase the productivity.

Similarly, in this research a survey is conducted in order to identify Agile practices and challenges. However the focus of this work is on distributed teams that work on a Global Software Development environment.

## 3 Method

The survey was conducted during the team's Innovation and Planning Iteration (IP), which is an event defined in SAFe's framework that is dedicated for Product Increment events, innovation activities, training and others. The exact period was from 11/April/2019 to 24/April/2019. During this period the survey was created using Microsoft Forms[2] and a link to the survey was provided by email to the team members. The survey remained open for 4 days and after that preliminary results were presented to the team during the IP Iteration Demo meeting.

The survey was composed of 18 questions, at which 17 were closed-ended questions and 1 was an open-ended question. Table 1 shows details about the types of questions in the survey:

---
[2] https://forms.office.com, last accessed 18-May-2019.

The data was then exported into Excel format, transformed as needed and upload to Python[3] analysis library, Pandas[4], so that data could be manipulated as needed and visual graphs could be generated accordingly. Other Python libraries were also used to generate different types of graphs.

It was received 32 responses, which represent around 19% of the population that the survey was sent to. Out of these 32 responses, 18 respondents answered the open-ended question which was the only question for which the answer was not mandatory among the 18 survey questions.

Table 1. Types of questions in the survey.

| Question Type | Allowed multiple answers | Answer Required | Likert | N. of Statements in Likert | N. of Options in Likert | Total by Type |
|---|---|---|---|---|---|---|
| Open-ended | NA | No | No | NA | NA | 1 |
| Close-ended | Yes | Yes | No | N/A | N/A | 3 |
| Close-ended | No | Yes | No | N/A | N/A | 8 |
| Close-ended | No | Yes | Yes | 10 | 5 | 1 |
| Close-ended | No | Yes | Yes | 1 | 5 | 5 |
| Total | | | | | | 18 |

Please note that the full list of the survey questions is in the appendix section.

## 4   Results

In this section, the results of each survey question is presented. The first question was to identify the role of the respondents. Since this survey was anonymous, those roles that have only one or two professionals were not explicitly listed, hence these are aggregated as 'Others'. As shown in Figure 1 the majority of the respondents were developers (14), followed by Quality Assurance - Tester (5), Technical Lead and Software Analyst (4 each), Development Manager and Scrum Master (2 each) and Other (1). No Architects and no Product Owners responded to the survey.

Question 2 was to identify how many years of experience in Agile Software Development the respondents have. As shown in Figure 2, it was found that 12 professionals have from 1 to 3 years of experience, 12 have 4 to 7 years of experience, 7 have more than 8 years of experience and 1 respondent has less than 1 year of experience.

In question 3, professionals were asked to select all Agile Methodologies they have experience with. In Figure 3 it is possible to see that Scrum is the most known framework by these professionals, followed by SAFe, which seems appropriate given the fact that SAFe is the framework used by the company as explained previously.

---

[3] https://www.python.org/, last accessed 18-May-2019.
[4] https://pandas.pydata.org/, last accessed 18-May-2019.

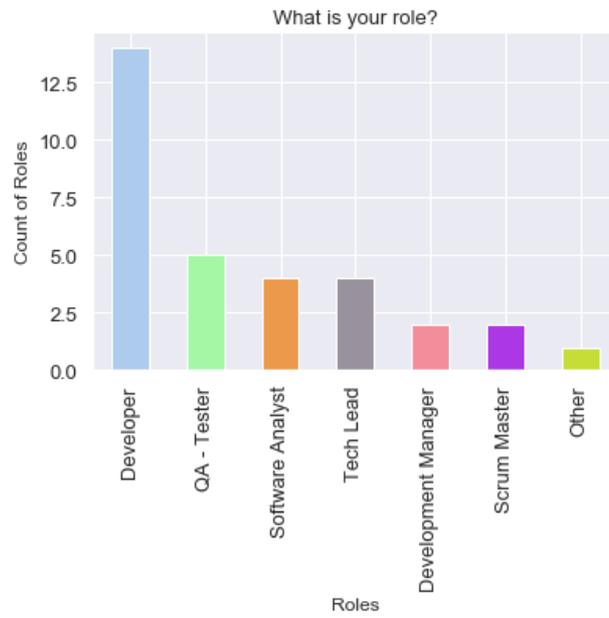

**Fig. 1.** Q1 - Roles distribution.

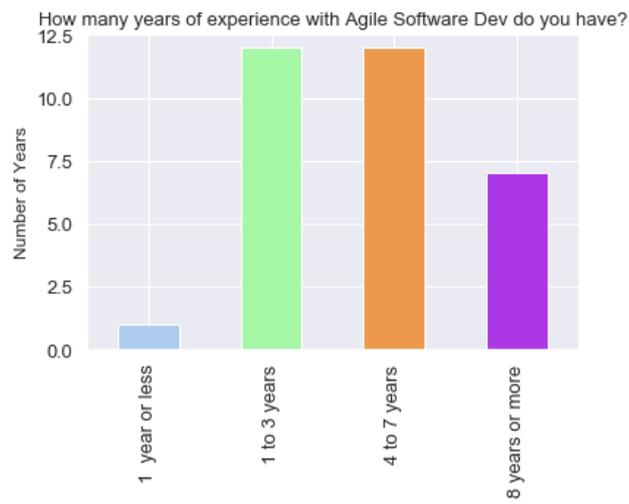

**Fig. 2.** Q2 - Years of experience in Agile Methodology.

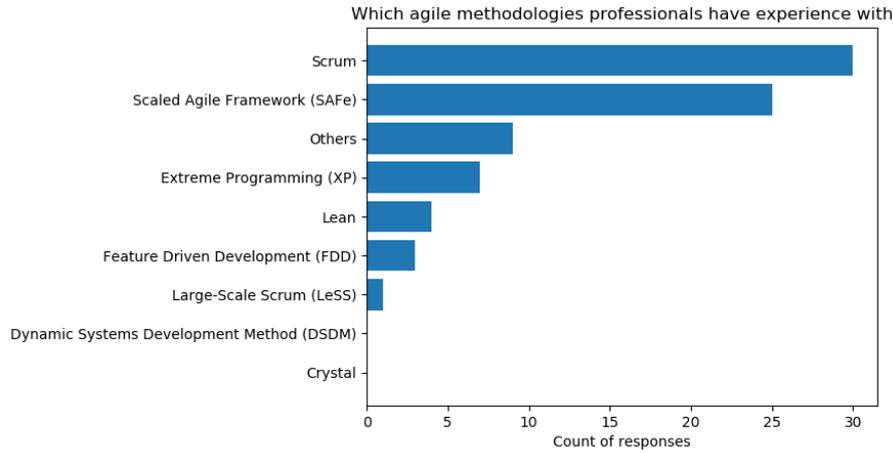

**Fig. 3.** Q3 - Agile Methodologies which professionals had experience with.

In question 4, participants were asked if they have already taken any training on any Agile methodology and it was found that 75% of the participants have already taken training on Agile methodology while 25% have not taken any training. Based on this result, the company could take actions to provide training courses to those who have not taken any training yet.

Question 5 presented a likert scale question, in which participants were asked to assess their familiarity with Agile methodologies in a scale of extremely familiarized, very familiarized, familiarized, not so familiarized or not familiarized at all. Figure 4 shows the results of their own assessment on this topic. In general, most participants feel they are either very familiarized or familiarized with Agile methodologies.

Question 6 presents another likert scale question, but this time participants were asked to assess their familiarity with SAFe. The results show that their familiarity decreased when compared to the previous question which was more generic as opposed to a specific framework as in question 6. However it is possible to see that most participants, 69% in total feel they are familiarized with SAFe.

In question 7, 10 different Agile practices and terms were selected and participants were asked to scale their confidence level on each of the selected practices and terms. Figure 6 shows the results in percentages per level of confidence. It is possible to see that a representative percentage of participants are not confident with a few practices, for example: 22% of the participants are not confident and 6% are not confident at all with Behaviour Driven Development practice, 25% are not confident and 3% are not confident at all with Test Driven Development practice, 22% are not confident and 6% are not confident at all with Pair Programming practice, 25% are not confident and 6% are not confident at all with Refactoring. With these results is it also possible to see that there are participants that feel extremely confident with some of these practice, perhaps that can indicate that knowledge transfer among the team members can increase the level of confidence to those who do not feel confident.

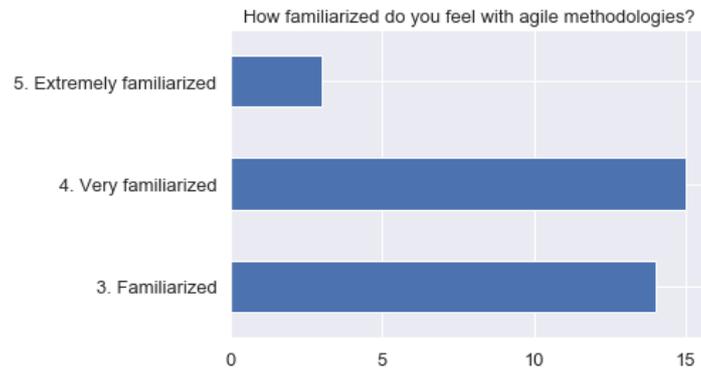

**Fig. 4.** Q5 - Familiarity with Agile Methodologies.

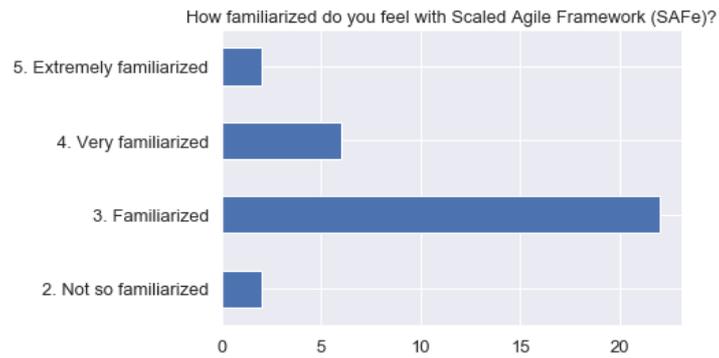

**Fig. 5.** Q4 - Familiarity with SAFe.

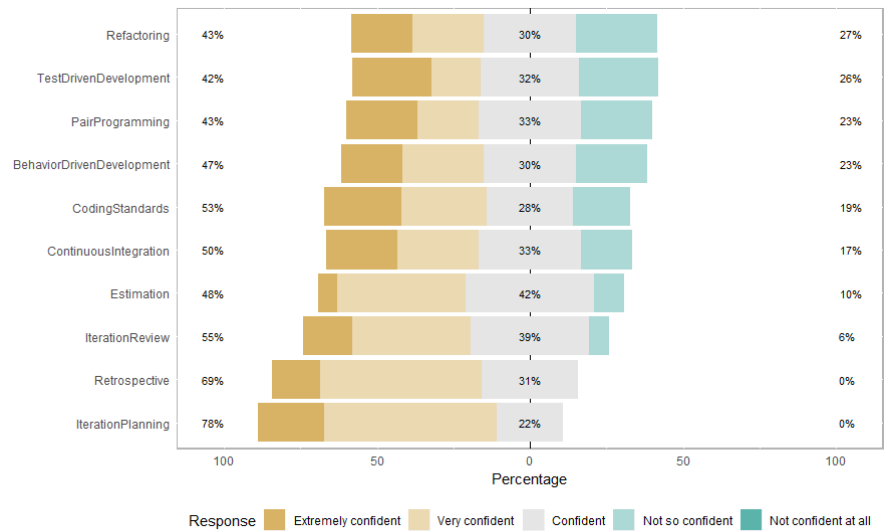

**Fig. 6.** Q7 - Confidence with Agile Practices or Terms.

Question 8 to 12 are all related to types of communication used by the participants and their evaluation of efficiency to some of these communication types. Figure 7 shows that e-mails and Skype chats are the types of communication most prioritized by these professionals, followed by 'Face-to-face, whenever possible' (22) and phone calls (19). 27 out of 32 participants selected email and Skype chats are their most prioritized type of communication. Only 1 participant selected video calls.

Figure 8 shows the biggest impediments for not communicating more via phone, face-to-face or via video calls. Time-zone constraints and agenda conflicts are the main causes, representing a total of 35% each.

Figure 9 shows how participants evaluate the efficiency of communication via e-mail, Skype chat and phone calls. It is possible to see that phone calls are considered the most efficient type of communication, followed by Skype chat and emails being the least efficient.

In question 13, participants were asked to respond how often they discuss project related items with the Product Owners (POs) or request feedback on features or stories, based on the fact that the 4th Agile principal, described in the Agile Manifesto says: 'Business people and developers must work together daily throughout the project'. Only 25% of the participants responded that they have daily communication with the Product Owner, 47% responded 'Once or twice per iteration', 19% responded 'Every other iteration' and 9% only during the Product Increment planning, which occurs every 3 months.

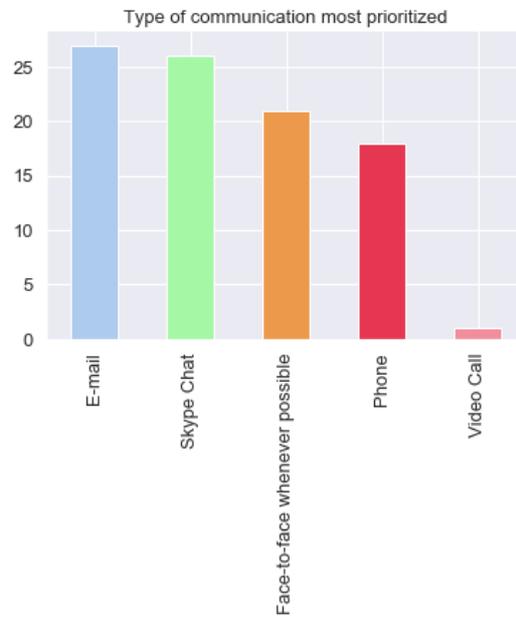

**Fig. 7.** Q8 - Types of communication prioritized.

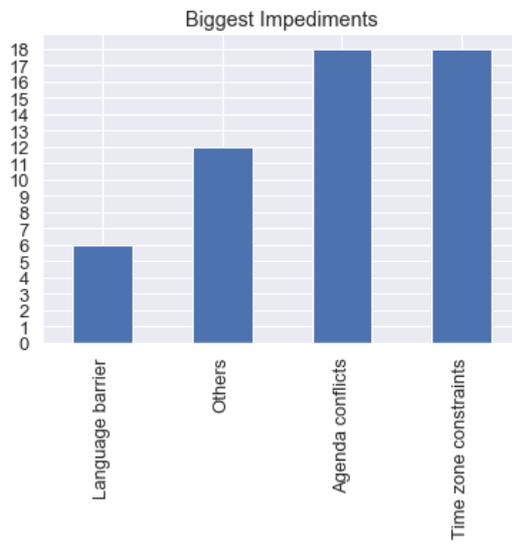

**Fig. 8.** Q9 - Impediments for not having more phone, face-to-face or video calls.

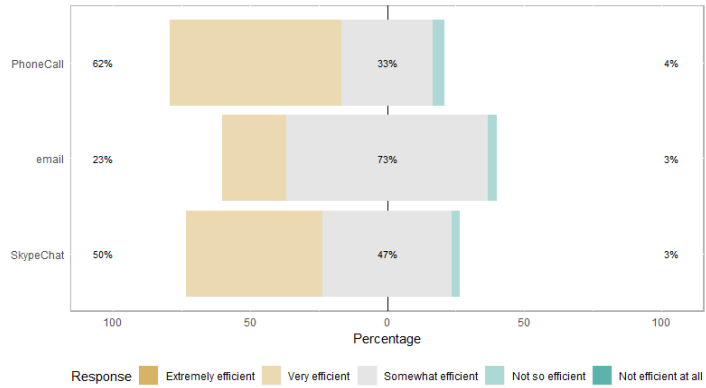

**Fig. 9.** Q10 - Evaluation of communication efficiency per type.

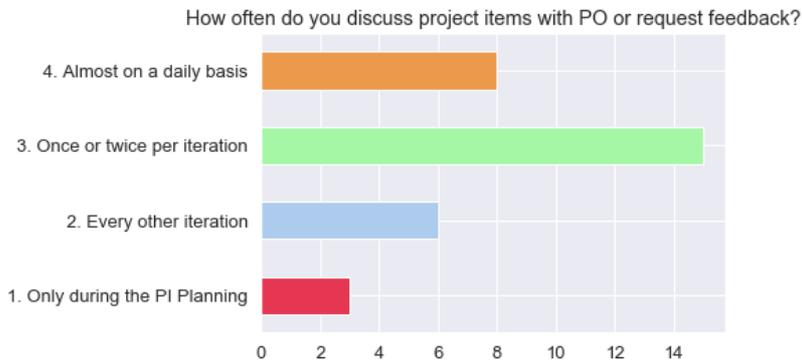

**Fig. 10.** Q13 - How often teams discuss project items with PO or request feedback.

Questions 14 and 15 were related to retrospective meetings. The results show that 97% of the respondents have retrospective meeting once per iteration and 91% said that retrospective meeting are resulting in actionable items to bring improvements, which is aligned with Agile principle 12 which says: 'At regular intervals, the team reflects on how to become more effective, then tunes and adjusts its behavior accordingly'.

In question 16, participants were asked if and how they were planning their capacity according to the team's velocity. The team's velocity in the company is measured in story points and to track team's capacity, a sum of story points that each team each team member can delivery for each iteration is made. Relative sizing estimation is used to measure story points, typically through pointing poker technique. The results show that 44% of the responds said their team's capacity is usually at 100% and 41% are usually at 80%.

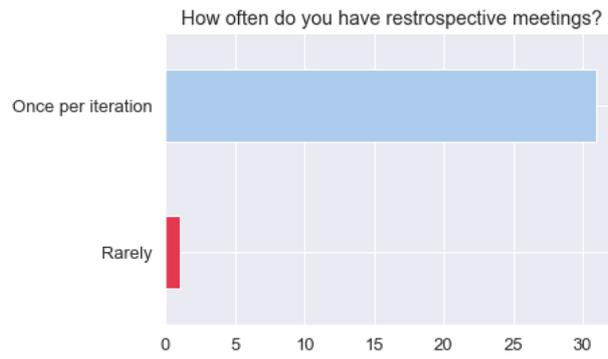

**Fig. 11.** Q14 - How often teams have retrospectives.

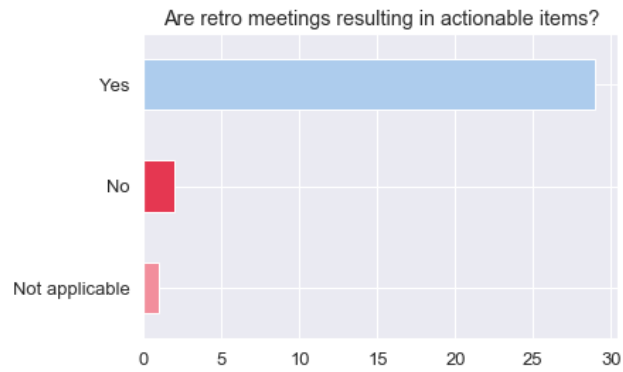

**Fig. 12.** Q15 - Are retrospectives resulting in actions/improvements?

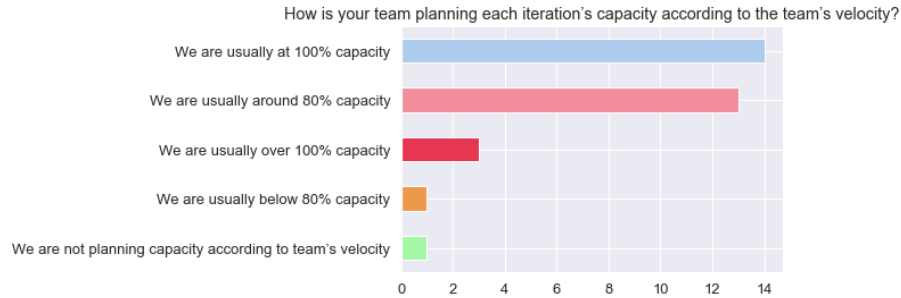

**Fig. 13.** Q16 - Are teams planning capacity based on velocity?

The last closed-ended question was related to how these professionals would control/track budget in an Agile project. The first SAFe principle is 'Take an economic view' and as per SAFe's guidelines, economics should inform and drive decisions at all levels, from Portfolio to Development Teams [5], therefore it is important that every team member has an idea of how to control budgets in an Agile project. Figure 14 shows that 63% of the respondents were not sure how to control budget, 22% responded that it would be 'Through planned and defined budget to cover the life cycle of the project', 9% responded 'Through incremental budget aligned in each phase' and 6% responded 'Through initial budget to cover MVP and the remaining budget to be discussed depending on MVP results'.

The last question was an open-ended question. Participants were asked what is/are the main challenge(s) of running an Agile development project with remote teams. Since this was an open-ended question, it was decided to generated a World Cloud graph, which is a visual representation of text data and the importance of each word is represented its size in the graph and based on the number of times these words were mentioned on the text data. In Figure 15 it become clear that communication is considered one of the main challenges raised by the participants.

Regarding the hypothesis raised in this study, which aimed to identify whether professionals with more experience in agile methodologies prioritize synchronous or asynchronous communication, Figure 16 shows a slightly higher correlation between years of experience and Skype chat communication if compared to other types of communication, although there is no strong correlation with any specific type of communication. There is strong correlation between types of communication prioritized, for example those who tend to prioritize Skype chat would also prioritize e-mail, those who tend to prioritize face-to-face communication would also prioritize phone calls.

**Fig. 14.** Q17 - How teams believe budget are controlled in Agile projects.

**Fig. 15.** Q18 - Word cloud with main challenges in running an Agile development project with remote teams.

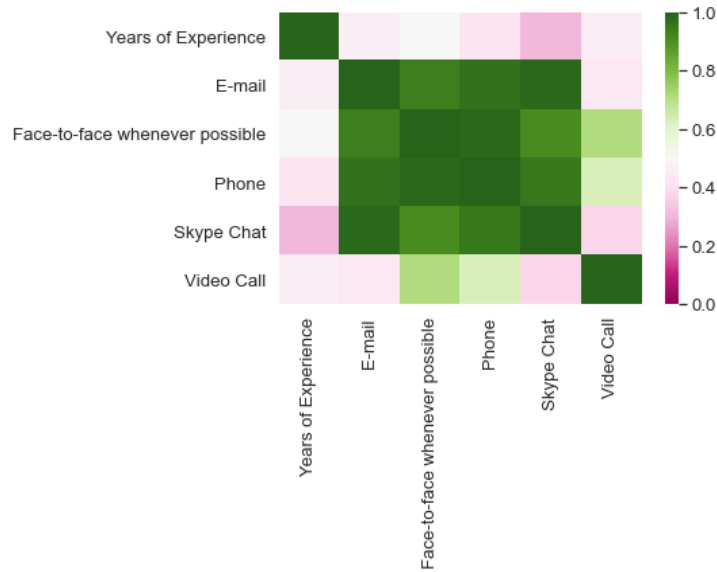

**Fig. 16.** Years of experience correlation with types of communication.

## 5    Conclusion

With the results of this study, it is clear that communication is one of the main challenges in running Agile projects in Global Software Development. Also, it was possible to confirm that there is no strong correlation between years of experience in Agile Software Development with types of communication prioritized. Results also showed that phone calls are considered to be the most efficient type of communication in Global Software Development environment. Additionally, it was possible to see that professionals have different levels of confidence in Agile practices, knowledge transfers could help level set teams overall confidence and knowledge.

## A  Appendices

### A.1  Agile Survey

**Objective:** The objective of this survey is to assess the level of knowledge in agile practices of the professionals and how communication barriers are overcome.

1. What is your role?
   **Options:** Developer, Scrum Master, Product Owner, QA - Tester, Tech Lead, Software Analyst, Development Manager, Architect, Other
2. How many years of experience with Agile Software Development do you have?
   **Options:** 1 year or Less, 1 to 3 years, 4 to 7 years, 8 years or more
3. Which agile methodologies do you have experience with? (Select all that apply)
   **Options:** Scrum, Extreme Programming (XP), Lean, Crystal, Dynamic Systems Development Method (DSDM), Feature Driven Development (FDD), Scaled Agile Framework (SAFe), Large-Scale Scrum (LeSS),Others
4. Have you ever attended any training on any Agile Methodology?
   **Options:** Yes, No
5. How familiarized do you feel with agile methodologies?
   **Options:** Extremely familiarized, Very familiarized, Familiarized, Not so familiarized, not at all familiarized

6. How familiarized do you feel with Scaled Agile Framework (SAFe)?
   **Options:** Extremely familiarized, Very familiarized, Familiarized, Not so familiarized, not at all familiarized
7. How would you classify your degree of knowledge in each Agile Practice/Term?
   **Options:** Extremely confident, Very confident, Confident Not so confident, Not at all confident Practices and Terms: Iteration Planning, Retrospective, Iteration Review, Behavior Driven Development, Test Driven Development, Coding Standards, Estimation, Pair programming, Continuous Integration, Refactoring
8. Which means of communication do you prioritize to interact with other scrum teams, product owners or other teams involved in the project delivery?
   **Options:** (Select all that apply) E-mail, Phone, Skype Chat, Video Call, Face-to-face whenever possible
9. If the answer to the previous question was e-mail or Skype chat, what is the biggest impediment for having more phone, face-to-face or video calls communication?
   **Options:** (Select all that apply) Language barrier, Time zone constraints, Agenda conflicts (For example: not being able to find available time in the person's agenda to have a phone call), Others
10. How efficient would you classify communication via e-mail?
    **Options:** Extremely efficient, Very efficient, Somewhat efficient, Not so efficient, Not at all efficient
11. How efficient would you classify communication via Skype Chat?
    **Options:** Extremely efficient, Very efficient, Somewhat efficient, Not so efficient, Not at all efficient
12. How efficient would you classify communication via Phone Call?
    **Options:** Extremely efficient, Very efficient, Somewhat efficient, Not so efficient, Not at all efficient
13. How often do you discuss project related items or request feedback on features/stories developed with your Product Owner?
    **Options:** Almost on a daily basis, Once or twice per iteration, Every other iteration, Only during the PI Planning
14. How often do you have retrospective meetings with your scrum team?
    **Options:** Once per iteration, Once a month, Rarely, Never
15. Are your retrospective meetings resulting in actionable items to bring improvements? If your team never has retrospective meetings, please select 'Not Applicable'
    **Options:** Yes, No, Not applicable
16. How is your team planning each iteration's capacity according to the team's velocity?
    **Options:** We are usually over 100% capacity, We are usually at 100% capacity, We are usually around 80% capacity, We are usually below 80% capacity, We are not planning capacity according to team's velocity
17. How would you control/track the costs/budgets of an Agile project?
    **Options:** Through planned and defined budget to cover the life cycle of the project, Through initial budget to cover MVP and the remaining budget

  to be discussed depending on MVP results, Through incremental budget aligned in each phase, Not sure
18. In your opinion, what is/are the main challenge(s) of running an Agile development project with remote teams? This is a open ended question and response on this is optional